\documentclass[reprint, superscriptaddress, amsmath, amssymb, aps, prl, twocolumn]{revtex4-2}

\usepackage[normalem]{ulem}
\usepackage{braket}
\usepackage{graphicx}  
\usepackage{epstopdf}
\usepackage{dcolumn}   
\usepackage{bm}        
\usepackage{amsmath}
\usepackage{amssymb}   
\usepackage{wrapfig}
\usepackage{array}
\usepackage{comment}
\usepackage{bbding}
\usepackage[bookmarks=false,linkcolor=blue,urlcolor=blue,colorlinks,citecolor=blue]{hyperref}

\usepackage{titlesec}
\titlespacing*{\section}
{0pt}{1.5ex}{0.8ex}  
\titlespacing*{\subsection}
{0pt}{1.2ex}{0.5ex}  
\begin{document}

\title{Metastability in Coexisting Competing Orders}


\author{Yasamin Masoumi Sefidkhani}
\email{masoumisefidkhani.y@northeastern.edu}
\affiliation{Department of Physics, Northeastern University, Boston, Massachusetts 02115, USA}

\author{Alberto de la Torre}
\affiliation{Department of Physics, Northeastern University, Boston, Massachusetts 02115, USA}
\affiliation{Quantum Materials and Sensing Institute, Northeastern University, Burlington, MA, 01803 USA}

\author{Gregory A. Fiete}
\affiliation{Department of Physics, Northeastern University, Boston, Massachusetts 02115, USA}
\affiliation{Quantum Materials and Sensing Institute, Northeastern University, Burlington, MA, 01803 USA}
\affiliation{Department of Physics, Massachusetts Institute of Technology, Cambridge, MA 02139, USA}

\begin{abstract}
The dynamical phase transition of a system with two coexisting competing order parameters is studied using the time-dependent-Ginzburg-Landau framework. The dynamics are induced by parameters capturing the physics of driving the system with an ultrafast laser pulse. A remarkable enhancement of the order parameter with a smaller mean-field value following the pump and the emergence of an induced metastable state is investigated through analytical and numerical studies. The effect of order parameter fluctuations on the exploration of the nonequilibrium free energy landscape reveals important information
about the impact of both thermal and nonthermal fluctuations on the dynamics of the metastable state.  Our results provide an interpretation of previously unexplained ultrafast experiments on superconductors with competing charge density wave order.  Our formalism is relevant across broad classes of out-of-equilibrium systems beyond the condensed matter context, such as the Kibble-Zurek cosmological model.
\end{abstract}

\maketitle


{\em Introduction}-Dynamical phase transitions (DPT)\cite{RevModPhys.49.435, BRAY199341, RevModPhys.83.863} occur between different states of a system under non-equilibrium conditions, typically driven by external time-dependent fields. Unlike phase transitions driven by changes in thermal variables such as pressure and temperature, DPT involve the temporal evolution of the system. DPT can be induced through non-thermal channels, such as an ultrafast laser quench, leading to transient and metastable states  \cite{metaa1, meta2, PhysRevResearch.3.033047, metadragon, NICOLIS200522}. Numerous studies have explored the rich phenomenology of ultrafast phase transitions \cite{RevModPhys.93.041002, PhysRevX.5.031039, PhysRevB.95.104507, RevModPhys.82.2731, doi:10.1073/pnas.1919451117, delaTorre2022,Claassen2017, Claassen2019,sokolowski2004ultrafast,tengdin2018critical,beaud2014time,RevModPhys.83.471,PhysRevB.90.100503,PhysRevB.98.224507,PhysRevB.98.020506,PhysRevLett.121.267003,PhysRevB.100.104507,PhysRevLett.125.267602}; however, they remain not fully understood.

There is a longstanding interest in competing and coexisting orders in materials \cite{mcmahon2020orbital,doi:10.1073/pnas.1612849113,PhysRevB.92.224502,Choi2020,Wu2011,PhysRevLett.101.207002,PhysRevLett.108.247002,PhysRevLett.107.237001,PhysRevResearch.5.L012032,PhysRevLett.101.207002,PhysRevB.66.094501,PhysRevB.109.064310}. Among the most interesting phenomena are DPT between different orders present in materials that exhibit high-temperature superconductivity (HSC). Materials exhibiting HSC are strongly correlated; therefore, there is a complex and intricate interplay between the different order parameters present in these systems. Often, the interactions between order parameters can lead to coexisting and competing orders such as charge density waves (CDW), spin density waves (SDW), and other magnetic orders alongside the HSC order \cite{RevModPhys.87.457, PhysRevB.103.195150, coexist, doi:10.7566/JPSJ.90.111005}. 

In condensed matter and materials physics, the advancements in pump-probe experiments have revealed novel phase transitions in systems with two or more order parameters that are either competing or intertwined. By pumping such systems with an ultra-fast laser, one can weaken the dominant order or even fully suppress it. Recent experiments \cite{Wandel,doi:10.1126/sciadv.abk0832,mcmahon2020orbital,PhysRevLett.113.107002} have investigated the interplay of CDW order with superconductivity (SC). Interestingly, after the pump the SC order is almost entirely suppressed \cite{Wandel,doi:10.1126/sciadv.abk0832} but the correlation length of the CDW order is increased beyond the thermal value \cite{Wandel} indicating a more robust order than prior to the pump. Moreover, there are other materials in which an ultrafast laser pump can drive dynamical phase transitions between different magnetic orders \cite{PhysRevB.106.165104,Li2022,doi:10.34133/ultrafastscience.0007}.  Even beyond the materials context discussed above, across all areas of physics it remains a general and broadly relevant question of how competing orders respond to out-of-equilibrium conditions \cite{RevModPhys.87.457,10.21468/SciPostPhys.8.5.074,proukakis2013quantum,PhysRevD.43.375}. 

\begin{figure}[t]
    \centering
\includegraphics[width=0.48\textwidth]{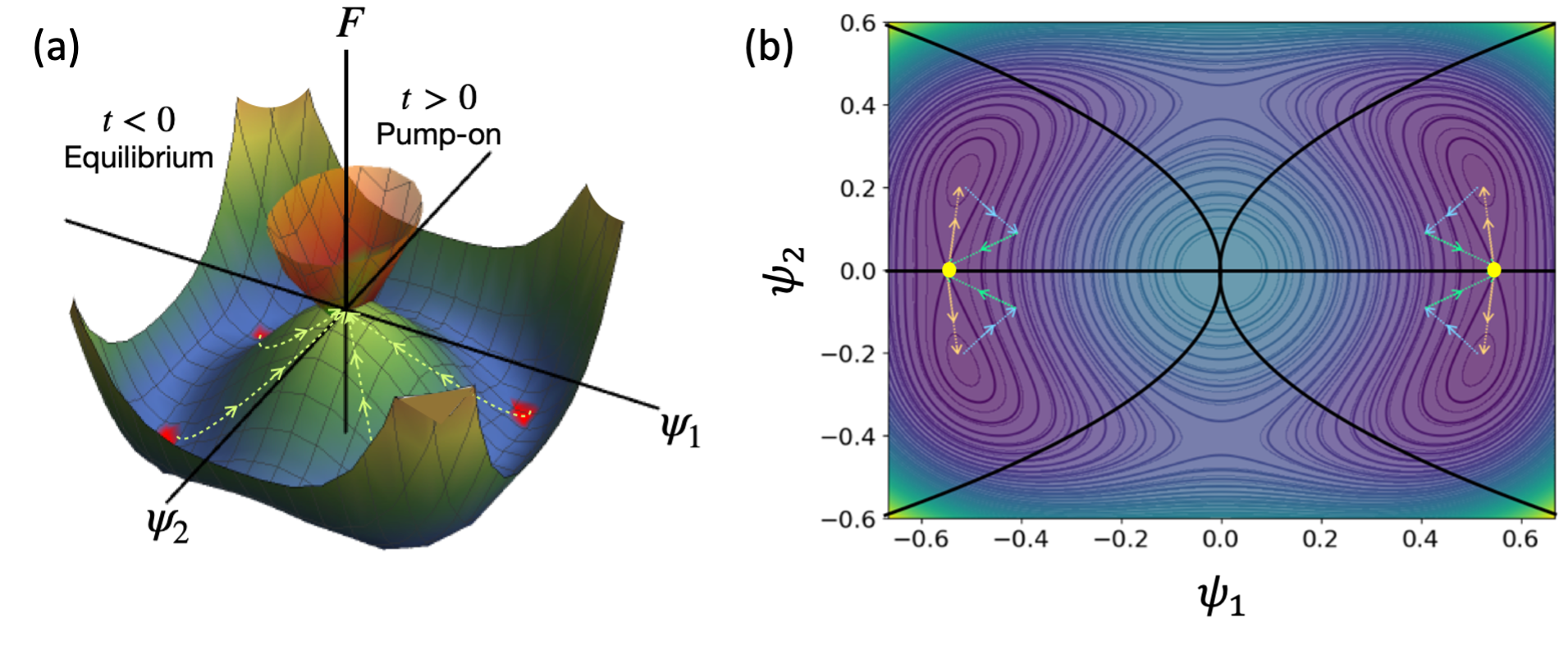}
    \caption{\footnotesize{(a) Free energy landscape at equilibrium with four global minima for two order parameters $\psi_1$, $\psi_2$. When the pump is on, the energy landscape changes to the orange parabola (only one minimum at the origin). Arrows indicate the population migration of each order parameter toward the origin. (b) Basins of attractions for the two order parameters in a contour plot of the free energy landscape. Darker purple colors indicate the minima. Solid black lines separate regions of basin attraction. Blue dotted lines show how $\psi_1$, $\psi_2$ decrease as the energy of the system increases in the pump-on regime. Green dotted lines show the dynamics after the laser is turned off, leading to an increase in the primary order and the suppression of the secondary order. The system ends up on a saddle point (yellow dots). The system remains in this metastable state for some time and slowly relaxes to the equilibrium shown by the orange dotted lines.}}
    \label{fig1}
\end{figure}

Extensive theoretical studies have explored systems with a single order parameter using time-dependent Ginzburg-Landau (TDGL) theory \cite{Mazenko1985, PhysRevLett.110.136404, Demler2020}, while more recently, the transient dynamics of systems with two strongly competing order parameters have been investigated \cite{Millis2020}. In this work, we consider and analyze how the amplitudes, spatial fluctuations, and correlation lengths of two coexisting order parameters initially in equilibrium evolve in time after a quenching by an ultrafast laser. We go beyond previous studies by including cross-correlation of the two orders, which were previously neglected and can account for nonthermal effects. Our model predicts the existence of dynamical trapping, a regime where the weaker order dominates over the one with a larger equilibrium amplitude, and the global system explores a local minimum of the free energy functional. We provide an expression for calculating the metastability lifetime of the dynamically trapped order parameters. When applied to a system exhibiting HSC and CDW coexistence, our theory explains previously unresolved experimental results \cite{Wandel,doi:10.1126/sciadv.abk0832}. However, our formalism is general and relevant to physical systems across all areas of physics, including the Kibble-Zurek cosmological model \cite{Zurek1985,ZUREK1996177,PhysRevE.81.050101,Chandrankibble}.

{\em Model-}Using the TDGL framework, we analyze thermal and non-thermal processes in a system with two scalar order parameters, $\psi_1(\vec{r},t)$(CDW) and $\psi_2(\vec{r},t)$(SC), which are spatially and temporally dependent.  Here $\vec r$ is the position and $t$ is the time. We focus on weakly competing orders where one order is more dominant, determined by the global minimum of the free energy functional \hyperref[fig1]{Fig.~\ref*{fig1}a}. 

If the laser couples to $\psi_2(\vec{r},t)$, SC with slower relaxation, the system can be trapped in a state where the primary order (i.e., the one with the largest transition temperature) is enhanced [$\psi_1(\vec{r},t)$, CDW], and the secondary order is nearly suppressed \hyperref[fig1]{Fig.~\ref*{fig1}b}. Our theory assumes a bath (e.g., dynamical phonons), making this an open system falling under Model A as proposed by Hohenberg and Halperin \cite{RevModPhys.49.435,Demler2020,PhysRevB.109.064310,delaTorre2022}.

The two order parameters in two spatial dimensions obey
\begin{equation}
    \frac{\partial \psi_i(\mathbf{r},t)}{\partial t}=-\Gamma_i \frac{\delta F}{\delta \psi_i(\mathbf{r},t)}+\eta_i(\mathbf{r},t),
    \label{eq1}
\end{equation}
where the energy functional $F$ with energy landscape shown in \hyperref[fig1]{Fig.~\ref*{fig1}}, is defined as,
\begin{equation}
    \begin{split}
        F = & \int d^2\mathbf{r}\left(f_1+ f_2 +c |\psi_1|^2|\psi_2|^2+\frac{L}{2}(\mathbf{\nabla}\times\mathbf{A})^2\right),
        \\  f_{i} = & \frac{1}{2}\alpha_i(t)|\psi_i|^2+\frac{1}{2}b_i |D_\mu \psi_i|^2+u_i |\psi_i|^4.
    \end{split}
    \label{eq2}
\end{equation}

The order parameters are complex. However, the phase of the two order parameters is taken to be spatially uniform and set to zero since we are interested in the thermodynamic properties of the system, which are determined by the amplitude variations \cite{doi:10.1080/00018732.2016.1194044,Yusupov2010}. In our notation, the energy functional density and order parameters are dimensionless. We take $\alpha_i(t<0)=-\alpha_i^{eq}<0$ to ensure a stable energy landscape with a nonzero minimum at equilibrium for both orders with $0<c<2\sqrt{u_1 u_2}$ (see SM)--the order parameters can coexist but also compete. 

At equilibrium, the system has a well-defined uniform temperature $T$. However, in the pump-on regime, we take the mass term, $\psi_i^2$, to have a temporal dependence that onsets the dynamics along with the field of the laser. We assume that the laser only changes the mass term of the order with a lower transition temperature, $\psi_2$. The evolution of the mass term is similar to Ref.\cite{PhysRevLett.113.026401},
\begin{equation}
\alpha_2(t) =
\begin{cases} 
-\alpha_2^{eq}<0 & \text{if } t < 0 \\
\alpha_2'>0 & \text{if } 0 < t < \tau_0 \\
(\alpha_2' + \alpha_2^{eq}) e^{-\frac{t-\tau_0}{t_c}} -\alpha_2^{eq} & \text{if } t > \tau_0
\end{cases}.
\label{eq4}
\end{equation}
This term $\alpha_2(t)$ is negative at equilibrium and the system exists in a minimized finite energy state. When the laser is on, the sign and magnitude of this term change, and the system no longer has a non-zero order parameter $\psi_2$ at the energy minimum \hyperref[fig1]{(Fig.~\ref*{fig1}a)}. When the laser is turned off the mass term relaxes back to its equilibrium value exponentially fast. The mass term of $\psi_1$ remains unchanged from equilibrium with $\alpha_1(t)=-\alpha_1^{eq}<0$.

We take the laser pump intensity, $I(t) \propto|\mathbf{A}(t)|^2$, to have temporal Gaussian widening 
which depends on the laser wave number $k_0$, the speed of light $c$, and the time interval when the laser is on $\tau_0$, through \cite{Esarey:95}, 
\begin{equation}
I(t)=I_0\left(1+4\left(\frac{t}{k_0 c \tau_0^2}\right)^2\right) e^{-2\left(\frac{t}{\tau_0}\right)^2}.
\label{eq5}    
\end{equation}

It is important to note that even though the mass term evolution proceeds in a non-thermal way through the last term in Eq.\eqref{eq1}, [$\eta_i(\mathbf{r},t),$ Gaussian noise with $\langle \eta_i(\mathbf{r},t)\rangle=0$] it ensures thermal equilibrium at later times by following the fluctuation-dissipation theorem \cite{Fluctuation} as $\langle\eta_i(\mathbf{r},t)\eta_j(\mathbf{r}',t')\rangle=\Gamma_i T \delta_{i,j}\delta(t-t')\delta(\mathbf{r}-\mathbf{r}')$.

We expand the order parameter $\psi_j(\mathbf{r})=\bar{\psi}_j+ \sum_{\mathbf{k}\neq0}\delta \psi_{j,\mathbf{k}}e^{i \mathbf{k}\cdot \mathbf{r}}$ to take into account its spatial variations about the mean-field value, $\bar{\psi}_j$. The spatial fluctuations $\delta \psi_{j,\mathbf{k}}$ are considered small compared to $\bar{\psi}_j$, according to the Ginzburg criterion since we assume that at equilibrium the system is sufficiently far from the critical temperature $T_c$. The connected correlation functions are defined as $C_{\mathbf{q}}^{ij}=\langle \psi^i_{\mathbf{q}}\psi^j_{-\mathbf{q}}\rangle_c$ and can describe both thermal effects coming from the bath and nonthermal effects coming from the time-dependence of the free energy functional.

{\em Dynamics-}In this section, we will give a general solution to the early dynamics of the system. The TDGL equations without approximations are, 
\begin{equation}
    -\frac{1}{\Gamma_{i}}\partial_t \bar{\psi_i} = r_i\bar{\psi_i}+4 c\bar{\psi_j}S^{ij},
\label{eq:mean}
\end{equation}
with $i\neq j$, and $i,j\in \{1,2\}$,
\begin{equation}
    r_i=\alpha_i(t)+b_1 e_i^2 I(t)+4u_i(\bar{\psi_i}^2+3 S^{ii}) +2 c (\bar{\psi_j}^2+S^{jj}),
    \label{eq10}
\end{equation}
and $S^{ij}=\frac{1}{(2 \pi)^2}\int_0^{\Lambda}d^2\mathbf{q}C^{ij}_{\mathbf{q}}$, where $\Lambda$ is a UV cutoff. 
The correlation dynamics are determined by
\begin{eqnarray}
    -\frac{1}{2\Gamma_{i}}\partial_t C_{\mathbf{q}} ^{ii} = \beta_i C_{\mathbf{q}}^{ii}+4c(\bar{\psi}_i \bar{\psi}_j +S^{ij}) C_{\mathbf{q}}^{ij} -T_{v},\;\;\\ 
    \label{eq8}
    \partial_t C_{\mathbf{q}}^{ij}=-\Gamma_i\left(\beta_i C_{\mathbf{q}}^{ij}+4c(\bar{\psi}_i\bar{\psi}_j +S^{ij})C_{\mathbf{q}}^{jj} \right) + (i\longleftrightarrow j),\;\;
    \label{eq9}
\end{eqnarray}
where $\beta_i=r_i + 8 u_i\bar{\psi}_i^2 + b_i \mathbf{q}^2$, and $T_{v}= T/(E_c )$, $T_v$ is the temperature from fluctuation-dissipation theorem that is normalized to the condensation energy of the whole system, $E_c$. 

\begin{figure}[t]
    \centering
\includegraphics[width=1\linewidth]{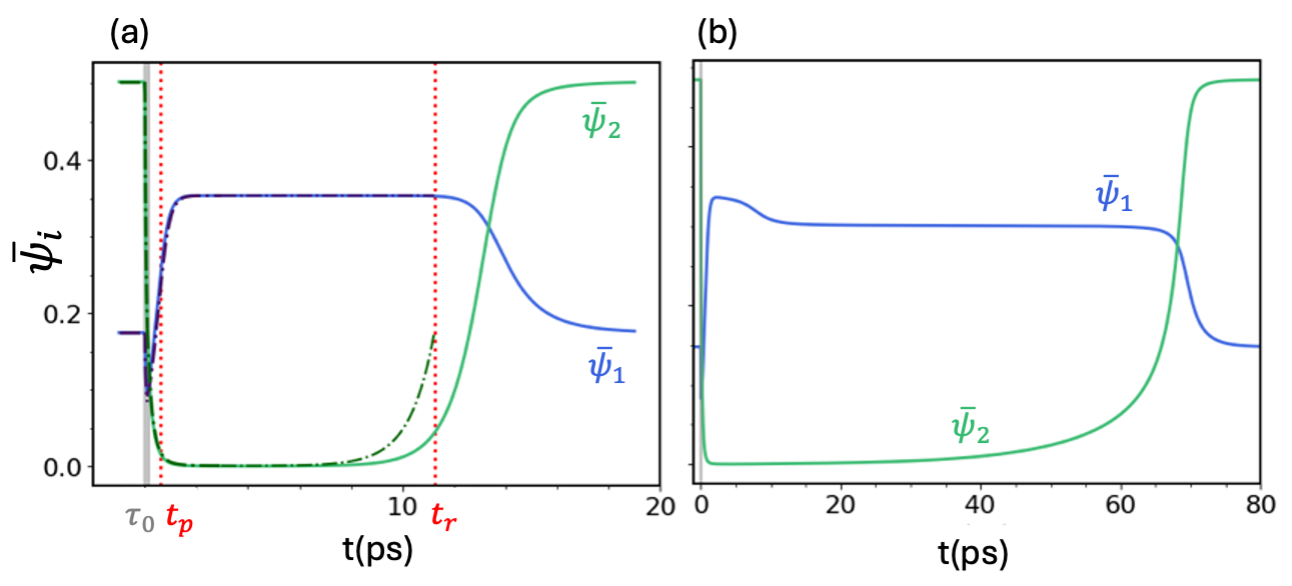}
\caption{\footnotesize{(a) The solid lines show the numerical solutions to Eq.\eqref{eq:mean} in the absence of fluctuations. The dashed lines are the analytical approximations of the equations. In the case of a strong quench, the analytical approximations describe the meanfield-only dynamics well before the system equilibrates. The metastability time, $t_r-t_p$, is around 11 ps, roughly $100 \tau_0$. (b) The solid lines are the numerical solutions to Eqs.\eqref{eq:mean}-\eqref{eq9}. The role of fluctuations in trapping the system for longer times is evident. Fluctuations allow the system to explore other regions of the energy landscape that are otherwise inaccessible. The presence of fluctuations of $\psi_2$ are so strong that they suppress $\bar{\psi}_2$ for some time.  The values used in the computation are: $\alpha_1=1.0$, $\alpha_2=1.1$, $\alpha_2'=3.5$, $b_1=7.0$, $b_2=2.0$, $u_1=2.0$, $u_2=1.0$, $c=1.5$, $\Gamma_1=2.5$ps\textsuperscript{-1} $\Gamma_2=1.5$ps\textsuperscript{-1}, $e^2I_0^2=0.7$, $T_v=0.07$, and $t_c=2.0$ ps. }}
    \label{fig3}
\end{figure}

In order to capture the physical picture of the dynamics of the system, we ignore the effect of fluctuations and only focus on Eq.\eqref{eq:mean}. If we assume the quench is strong enough for one to ignore all but the linear term in the meanfield-only version of Eq.\eqref{eq:mean}, then in the pump-on regime, $t\leq \tau_0$ (the thin grey area near $\tau_0$ in \hyperref[fig3]{Fig.~\ref*{fig3}a}), the solution becomes an exponential decay,
\begin{equation}
    \begin{split}
        \bar{\psi}_1(t) &= \bar{\psi}_1^0  e^ {-\frac{1}{4}\Gamma_1 b_1 I_0\tau_0 \zeta(t)},
        \\ \bar{\psi}_2(t)& = \bar{\psi}_2^0  e^ {-\Gamma_2\alpha'_2 t} e^ {-\Gamma_2 b_2 I_0 \tau_0\zeta(t)},
    \end{split}
    \label{eq11}
\end{equation}
with $\zeta(t)=\sqrt{2\pi}\left(1+ \frac{1}{\sigma^2}\right)\text{Erf}\left(\frac{ \sqrt{2}t} {\tau _0}\right)- 4\frac{1}{\sigma^2}\frac{t}{\tau_0} e^{-2(\frac{t}{\tau_0})^2}$ and  $\sigma=k_0 c\tau_0$. This result is similar to Ref.\cite{Millis2020}; however, both orders are non-zero and the quench is far from equilibrium without near-ergodic conditions \cite{Yusupov2010}. The results we now turn to were not reported previously.

For the pump-off regime at early times, $\tau_0\leq t$, one can calculate the solution to $\bar{\psi}_2$ very similar to the previous Eq.\eqref{eq11}, with the laser field being zero and the changing mass term as,
 \begin{equation}
    \bar{\psi}_2(t)= \bar{\psi}_2(\tau_0) e^{\Gamma_{2}\left(\alpha_2^{eq} (t-\tau_0) + (\alpha'_2+ \alpha_2^{eq})t_c (e^{-\frac{t-\tau_0}{tc}}-1)\right)}.
    \label{eq12}
\end{equation}
Therefore, $\bar{\psi}_1$ can be determined by including all the terms in the meanfield-only version of Eq.\eqref{eq:mean} as a solution to a second-order Bernoulli differential equation.

\begin{figure}[t]
    \centering
\includegraphics[width=1\linewidth]{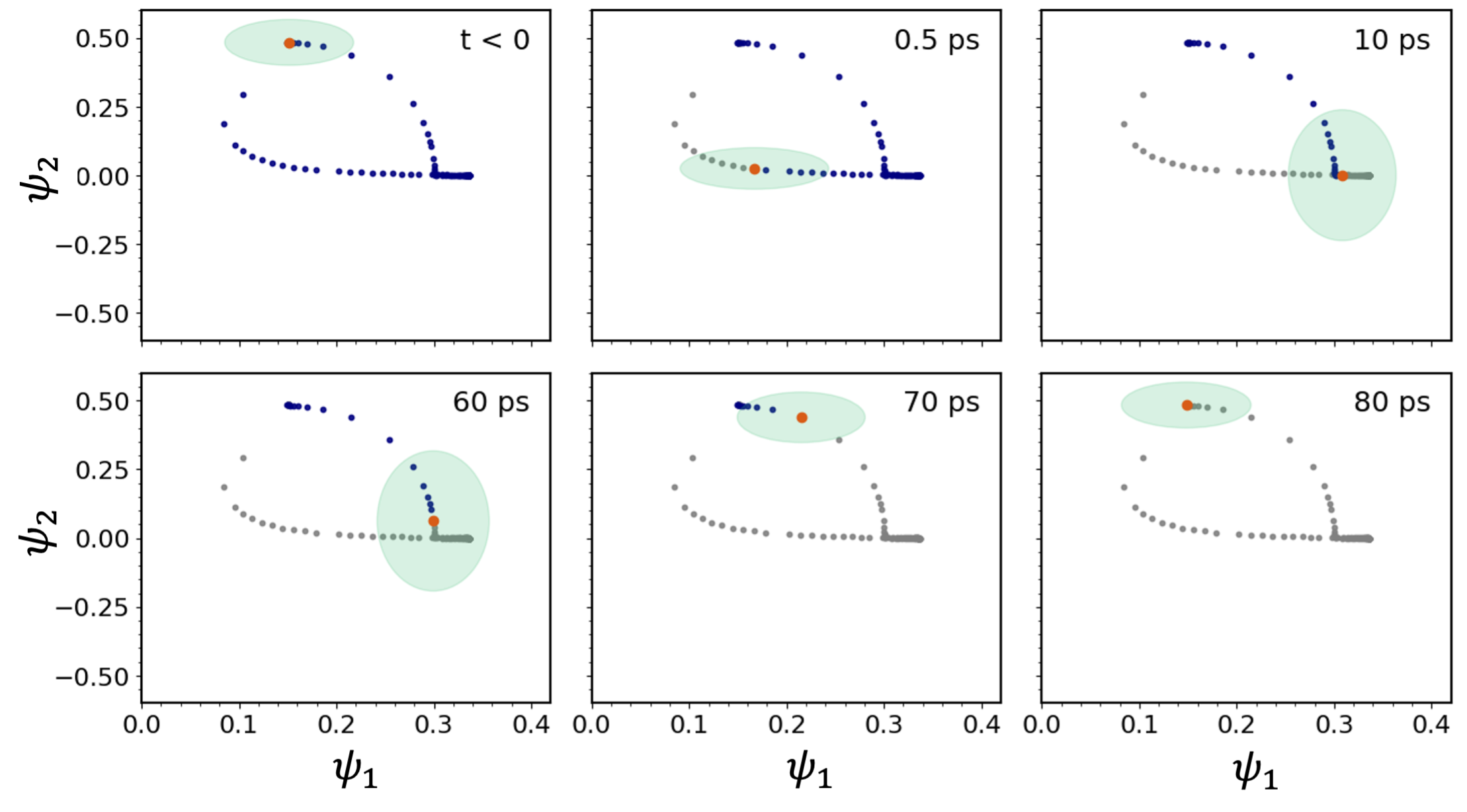}
\caption{\footnotesize{The state of the system is depicted at various snapshots in time. The orange point is the meanfield value and the green ellipse shows the fluctuation of each order. The dark blue dots show the future of the mean-field values and the grey ones the past. The laser pump-on regime is $0<t<0.1$ps.  At $t<0$ the system is at equilibrium with larger fluctuations for the order, $\psi_1$. At $t=0.5$ps, the system shows a full suppression of $\psi_2$ accompanied by a suppression of its fluctuations, which is the opposite of the dynamics of $\psi_1$ (where both are increased). At $t=10$ps, the system exhibits only an increase of $\psi_1$. At later times, $t=60$ps, one can see the metastability of the meanfield values and the increase of the fluctuations of $\psi_2$ for its revival. At $t=70$ps, the slow revival of $\psi_2$ is depicted. Later, at $t=80$ps the system returns to equilibrium. The values are the same as \hyperref[fig3]{Fig.~\ref*{fig3}}.}}
    \label{fig4}
\end{figure}

Given the initial conditions and Eq.\eqref{eq12}, we calculate the time $t_p$ at which $\psi_2(t_p)\ll\psi_2^0$ and $\psi_2(t_p)\ll\psi_1(t_p)$, as seen in \hyperref[fig3]{Fig.~\ref*{fig3}a}. At this part of the dynamics, the effect of the primary OP, the order that is more robust after the quench, on the secondary OP, the order that is suppressed after the quench, cannot be neglected. Thus, as an approximation, the competition term ($c \psi_1^2 \psi_2^2$) is ignored for the first OP. In contrast, all terms except the quartic term in the energy functional are included for the secondary OP. This leads to
\begin{equation}
    \bar{\psi}_1(t) = \sqrt{\frac{\alpha_1^{eq}}{4 u_1 -(4 u_1-\frac{\alpha_1^{eq}}{\bar{\psi}_{1,p}^2}) e^{-2 \Gamma_1 \alpha_1^{eq} (t-t_p)}}},
    \label{eq14}
\end{equation}
\begin{equation}
\bar{\psi}_2(t)=\bar{\psi}_{2,p}e^{\Phi(t)}\left( \frac{4 u_1 e^{2 \Gamma_1 \alpha_1(t-t_p)}-\frac{4 u_1 \bar{\psi}^2_{1,p}-\alpha_1}{\bar{\psi}^2_{1,p}}}{4 u_1-\frac{4 u_1 \bar{\psi}^2_{1,p}-\alpha_1}{\bar{\psi}^2_{1,p}}}\right)^{-\frac{\Gamma_2}{\Gamma_1}\cdot\frac{c}{4 u_1}}\!,
    \label{eq15}
\end{equation}
$\Phi(t)=\Gamma_2\left(t_c(\alpha'_2+\alpha_2)\left(e^{-\frac{t-\tau_0}{tc}}-e^{-\frac{t_p-\tau_0}{tc}}\right)+\alpha_2(t-t_p)\right)$, and $\psi_i(t_p)=\psi_{i,p}$. Eqs.\eqref{eq14} and \eqref{eq15} show $\bar{\psi}_2\approx 0$ with $\bar{\psi}_1^2(t)\rightarrow\frac{\alpha^{eq}_1}{4u_1}$ for a time after $t_p$ and less than $t_r$ \hyperref[fig3]{Fig.~\ref*{fig3}}a, which is an extremum point in the energy landscape and corresponds to a metastable state. We remark that since $-\frac{\alpha_2^2}{16 u_2}<-\frac{\alpha_1^2}{ 16u_1}$, the metastable state has the highest energy among all three extrema present in the system. By taking a look at the longtime behavior of Eq.\eqref{eq15}, one can see that there is an exponential growth present since $\frac{\alpha_2^2}{ u_2}>\frac{\alpha_1^2}{ u_1}$. The consequence of this behavior is that the small-$\bar{\psi}_2$ approximation breaks down after some time, $t_r$, as shown in Fig.~\ref{fig3}a and gives us an upper bound for the metastability lifetime. The cutoff for this approximation can be taken to be $\bar{\psi}_2(t_r)=\bar{\psi}_1^0$. The trapping time, $t_r-t_p$ (see \hyperref[fig3]{Fig.~\ref*{fig3}}a), is implicitly determined from
\begin{equation}
    \begin{aligned}
\frac{2 u_1}{\Gamma_2 c} \ln\frac{\bar{\psi}_{2,p}}{\bar{\psi}_1^0} &= \left(\alpha_1^{eq} - 2\frac{u_1}{c}\alpha_2^{eq}\right)(t_r - t_p) \\
&\quad - \frac{2 u_1}{c}t_c(\alpha_2' + \alpha_2^{eq}) e^{-\frac{t_p - \tau_0}{t_c}} \left(e^{-\frac{t_r - t_p}{t_c}} - 1\right).
\end{aligned}
    \label{eq:timemeta}
\end{equation}
{\em Metastability and Fluctuations-}
In \hyperref[fig4]{Fig.~\ref*{fig4}} we present 
snapshots in time of the fluctuations of the out-of-equilibrium order parameters around their meanfield value. By comparing Eq.\eqref{eq:mean}-\eqref{eq9} with their meanfield-only counterparts, one can see that the relaxation rates of $\psi_{1,2}$ are, in general, different. Assuming that the conditions are such to trap the system in a metastable phase--a strong quench, where one of the orders is almost fully suppressed accompanied by the enhancement of the other--the dynamics are nearly frozen. How long the system spends in the metastable state depends on the magnitude of the right-hand side of the TDGL equations, Eqs.\eqref{eq:mean}-\eqref{eq9}. Furthermore, since $C^{ii}_\mathbf{k}$ is positive, it will slow down the revival of the order, which leads to a longer metastable lifetime. These arguments explain the numerical results in \hyperref[fig3]{Fig.~\ref*{fig3}b}.

The fluctuations manifest in renormalizing the equilibrium values as well as the metastable ones. If one compares \hyperref[fig3]{Fig.~\ref*{fig3}a} and \hyperref[fig3]{Fig.~\ref*{fig3}b} one can see that in the case where fluctuations are included the system initially goes to the saddle point \hyperref[fig1]{Fig.~\ref*{fig1}b} and then relaxes to a point around the saddle point. This system can explore other parts of the energy landscape by tuning the magnitude of the fluctuations through the temperature of the bath. 
\begin{table}[b]
    \centering
    \caption{Different regimes of order parameter interactions.}
    \begin{tabular}{c c c}
        \hline
        \textbf{Condition} & \textbf{Regime} & \textbf{Order} \\ \hline \\[-4pt]
        $c<0$ & Cooperating& $\bar{\psi}_1,\bar{\psi}_2\neq 0$, $C^{ij}_\mathbf{k}>0$  \\[4pt]
        $0<c<2\sqrt{u_1 u_2}$ & Weakly competing & $\bar{\psi}_1,\bar{\psi}_2\neq 0$, $C^{ij}_\mathbf{k}<0$ \\[4pt]
        $2\sqrt{u_1 u_2}<c$& Strongly competing & $\bar{\psi}_1\neq 0$, $\bar{\psi}_2=0$, $C^{ij}_\mathbf{k}=0$\\[4pt]
        \hline
    \end{tabular}
    \label{tab:table_regimes}
\end{table}


{\em Cross Correlations-}Since, in our system, both order parameters have a nonzero expectation value, their cross-correlation $C_\mathbf{k}^{ij}$ is nonzero and is therefore important in the dynamics, as seen in Table~\ref{tab:table_regimes}. In the presence of this term, the increase of one order parameter can lead to a decrease of the other one and this makes the dynamics of both orders less robust.  Mathematically, if one looks at Eqs.\eqref{eq:mean}-\eqref{eq9}, this term appears in the meanfield equation and its contribution is negative. This leads to a smaller damping and a faster recovery of the system when trapped in a metastable phase. This can also be seen in the calculation of the correlation length of both orders. In the absence of this term, the correlation length increases significantly, implying the orders become more robust(See SM).

\begin{figure}[t]
    \centering
\includegraphics[width=1\linewidth]{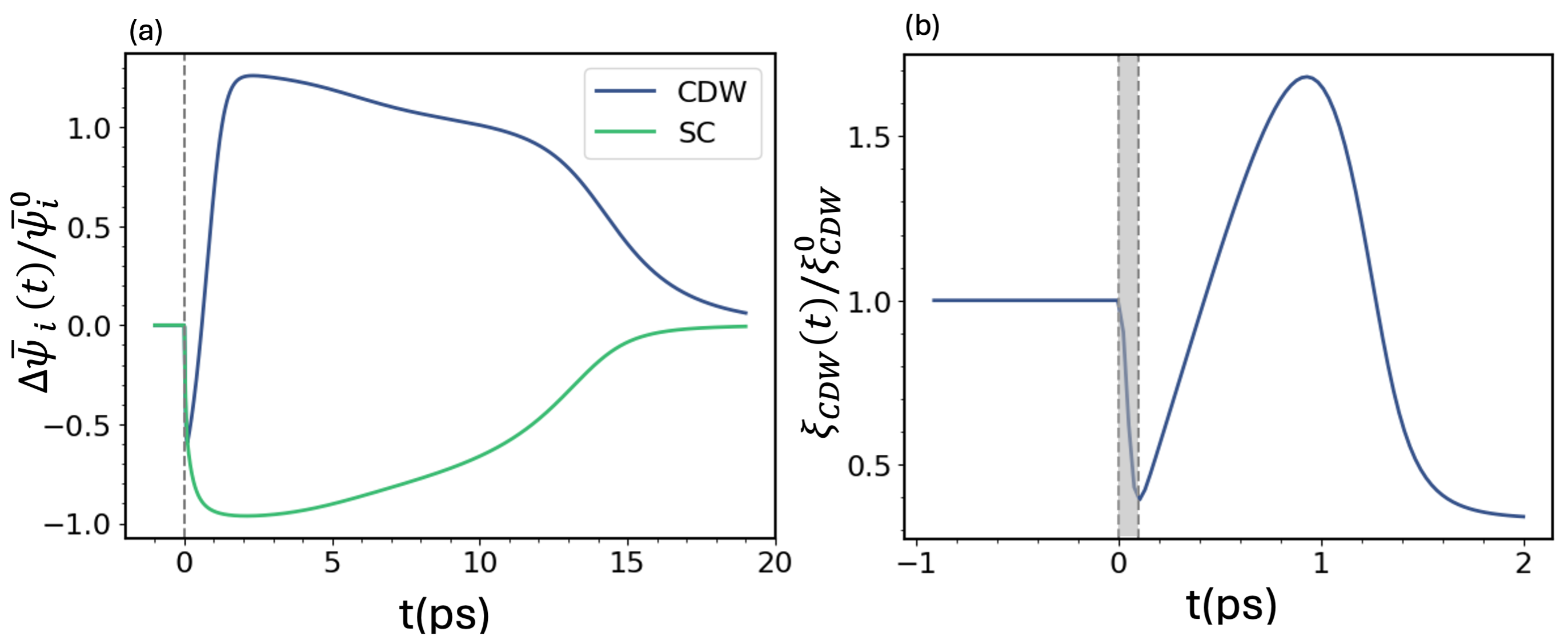}    \caption{\footnotesize{(a) The relative changes of the meanfield value of each order are plotted. After a pump, the orders are suppressed for a very short amount of time, and $\sim$1ps after the laser quench the CDW order increases with the SC order remaining suppressed. (b) The correlation length of the CDW order relative to equilibrium. The correlation length increases by around 70\% compared to its equilibrium value after about 1ps which leads to the enhancement of the order seen in panel (a). The values are $\alpha_\mathrm{CDW}=1.0$, $\alpha_\mathrm{SC}=1.1$, $\alpha_\mathrm{SC}'=3.5$, $b_\mathrm{CDW}=7.0$, $b_\mathrm{SC}=2.0$, $u_\mathrm{CDW}=2.0$, $u_\mathrm{SC}=1.0$, $c=1.5$, $\Gamma_\mathrm{CDW}=2.5$ps\textsuperscript{-1} $\Gamma_\mathrm{SC}=1.0$ps\textsuperscript{-1}, $e^2I_0^2=1.0$, $T_v=0.07$, and $t_c=1.0$ ps.}}
    \label{fig6}
\end{figure}

{\em Experiments-}An example of a physical system that exhibits two competing yet coexisting orders is the YBCO compound \cite{doi:10.1126/science.1223532}. There have been numerous studies demonstrating the coexistence of a charge density wave (CDW) order alongside a high-T\textsubscript{c} superconductivity below a critical temperature of around 60K in $\mathrm{YBa}_2\mathrm{Cu}_3\mathrm{O}_{6.67}$. The out-of-equilibrium aspect of these two orders has been studied recently and there have been intriguing behaviors exhibited through an ultrafast laser quench \cite{Wandel,doi:10.1126/sciadv.abk0832}. In these experiments, after shining light on the YBCO compound, a 120\% enhancement of the CDW amplitude with more than 90\% suppression of the SC order has been observed. The experiment also shows a fascinating increase in the correlation length of the CDW order. Our theory presented here, predicts these behaviors very well. As shown in \hyperref[fig6]{Fig.~\ref*{fig6}a} around 120\% increase in the meanfield value of the CDW is predicted accompanied by around 70\% increase in the correlation length of the charge density wave \hyperref[fig6]{Fig.~\ref*{fig6}b}. 

{\em Summary-}In this work we have used a time-dependent Landau-Ginzburg framework to study scenarios relevant to common classes of pump-probe experiments in which the pump strongly suppresses an order which then recovers in time after the laser pulse passes through the material.  We focused on the interesting case where two orders are present and competing with one another.  We have found: (i) Suppressing the weaker of the two orders can lead to an enhancement of the other order beyond its equilibrium value. (ii) An analytical formula estimating the the metastability lifetime of the enhanced primary order. (iii) The quantitative effects of thermal and non-thermal order parameter fluctuations on the metastability lifetime by direct numerical evaluation of the coupled orders.  (iv) The effect of cross correlations between orders on the metastability lifetimes.  Our work is directly relevant to materials with competing orders, such as superconductivity and charge density waves, and the formalism is applicable to broad classes of systems across all areas of physics.

{\em Acknowledgments-}We thank S. Chaudaury, M. Buchhold, A. Chandran, and M. Classen for their helpful discussions. We acknowledges funding from the National Science Foundation through DMR-2114825. G.A.F. acknowledges additional support  from the Alexander von Humboldt Foundation. 



%

\clearpage
\onecolumngrid
\begin{center}
    \textbf{\large Supplementary Material for: Metastability in Coexisting Competing Orders}\\
    \vspace{0.5cm}
    \textbf{Yasamin Masoumi Sefidkhani$^1$, Alberto de la Torre$^{1,2}$, Gregory A. Fiete$^{1,2,3}$}\\
    \vspace{0.5cm}
    \textit{$^1$Department of Physics, Northeastern University, Boston, Massachusetts 02115, USA}\\
    \textit{$^2$Quantum Materials and Sensing Institute, Northeastern University, Burlington, MA, 01803 USA}\\
    \textit{$^3$Department of Physics, Massachusetts Institute of Technology, Cambridge, MA 02139, USA}
\end{center}


\section{Equilibrium Free Energy}
\renewcommand{\theequation}{I.\arabic{equation}}
\setcounter{equation}{0}

By taking partial derivatives of the energy functional, we can find the energy minima. In order to simplify the notation we take $\alpha_i^{eq}=\alpha_i$

\begin{equation}
\begin{split}
\frac{\delta F}{\delta\bar{\psi}_2} = & -\alpha_2\bar{\psi}_2+4 u_2\bar{\psi}_2^3+2 c\bar{\psi}_1^2\bar{\psi}_2=0,  
\\ \frac{\delta F}{\delta\bar{\psi}_1} = & -\alpha_1\bar{\psi}_1+4 u_1\bar{\psi}_1^3+2 c\bar{\psi}_2^2\bar{\psi}_1=0.
\end{split}
\label{appa1}
\end{equation}
The solutions to the above are:
\begin{equation}
\begin{split}
\text{1.} \quad & \bar{\psi}_1=0, \quad \quad \bar{\psi}_2^2=\frac{\alpha_2}{4 u_2}, \\
\text{2.} \quad & \bar{\psi}_2=0, \quad \quad \bar{\psi}_1^2=\frac{\alpha_1}{4 u_1}, \\
\text{3.} \quad & \bar{\psi}_2^2= \frac{\alpha_1 c -2 u_1 \alpha_2}{2 c^2 -8 u_1 u_2}, \quad \quad \bar{\psi}_1^2=\frac{\alpha_2 c - 2 u_2 \alpha_1}{2 c^2 -8 u_1 u_2}.
\end{split}
\label{appa2}
\end{equation}
Let us assume case 1 has a smaller energy compared to case 2, $-\frac{\alpha_2^2}{16 u_2}<-\frac{\alpha_1^2}{ 16 u_1}$ and compare case 1 and 3. (Comparing case 2 and 3 with the inequality reversed produces the same result.) First, by calculating the Hessian matrix, we see that 1 and 2 are saddle points and case 3 is a local minimum. In order to have case 3 as the global energy minimum, $F(3)<F(1)$, 
\begin{equation}
    \frac{(\alpha_1 c -2 \alpha_2 u_1)^2}{16 u_1(c^2- 4 u_1 u_2)}<0.
    \label{appa3}
\end{equation}
From the denominator, one obtains $0<c<2\sqrt{u_1 u_2}$. 

\section{Derivation of the Equations of Motion}

\renewcommand{\theequation}{II.\arabic{equation}}
\setcounter{equation}{0}

From Eq.\eqref{eq1}, one can calculate the dynamics of both the meanfield values and the correlation functions. First, one needs to separate the fields into meanfield values and the small fluctuations as,
\begin{equation}
\psi_i(\mathbf{r},t)=\bar{\psi}_i+\delta\psi_i(\mathbf{r})=\bar{\psi}_i+\sum_{\mathbf{k}\neq 0}\delta\psi_{i,\mathbf{k}}e^{i\mathbf{k}\cdot \mathbf{r}}.
    \label{appb1}
\end{equation}
By looking at the expectation value of the spatially uniform part, $\mathbf{k}=0$, the meanfield dynamics is
\begin{equation}
\frac{\partial \bar{\psi}_i}{\partial t} =-\Gamma_i \left\langle \frac{\delta F}{\delta \psi_i}+ \eta_i(\mathbf{r},t)\right\rangle_{\mathbf{k}=0}.
    \label{appb2}
\end{equation}
The dynamics of the fluctuating part are very similar as seen by looking at the $\mathbf{k}$-component of the Fourier transform
\begin{equation}
\frac{\partial \delta\psi_{i,\mathbf{k}}}{\partial t} =-\Gamma_i \left( \frac{\delta F}{\delta \psi_i}\right)_\mathbf{k\neq 0} +\left( \eta_i(\mathbf{r},t)\right)_\mathbf{k \neq 0}.
    \label{appb3}
\end{equation}
The dynamical equation for the correlation functions, $C^{ii}_{\mathbf{k}}$ is calculated as,
\begin{equation}
\small
    2 \delta\psi_{i,-\mathbf{k}}\frac{\partial \delta\psi_{i,\mathbf{k}}}{\partial t}=-2\Gamma_i \delta\psi_{i,-\mathbf{k}}\left( \frac{\partial F}{\partial \psi_i}\right)_\mathbf{k} +2 \delta\psi_{i,-\mathbf{k}}\left( \eta_i(\mathbf{r},t)\right)_\mathbf{k}.
    \label{appb4}
\end{equation}

By taking an expectation value over all the possible states of the system the correlation function $\partial_t C^{ii}_{\mathbf{k}}$ can be calculated using Wick's theorem. Unlike most cases since the fields present here are not centered around zero and have a nonzero expectation value, the cross-correlation, $C^{ij}_\mathbf{k}=\langle\psi_\mathbf{-k}^i\psi_\mathbf{k}^j\rangle_c\neq 0$. Therefore, the extra terms coming from this contraction cannot be omitted. This leads to $\partial_t C^{ij}_{\mathbf{k}}$, with nonnegligible dynamics that is calculated in a similar manner.


\section{Analytical Solutions}
\renewcommand{\theequation}{III.\arabic{equation}}
\setcounter{equation}{0}
\setcounter{subsection}{0}
\subsection{pump-on regime}

Due to the complexity of the equations, to capture the essence of the mean-field dynamics, we will ignore all the correlations and look only at the mean-field equations,
\begin{equation}
\begin{split}
    -\frac{1}{\Gamma_{1}}\partial_t \bar{\psi}_1=&(-\alpha_1 + b_1  e^2 \vec{A}(t)^2 +4 u_1 \bar{\psi}_1^2+ 2 c \bar{\psi}_2^2)\bar{\psi}_1,
   \\ -\frac{1}{\Gamma_2}\partial_t \bar{\psi}_2=&(\alpha_2(t)+ 4 b_2  e^2 \vec{A}(t)^2+4 u_2 \bar{\psi}_2^2+ 2 c \bar{\psi}_1^2)\bar{\psi}_2.
    \end{split}
    \label{appc1}
\end{equation}

When the laser is on, $\vec{A}\neq 0$ and the system starts its dynamics. When the pump is on $\vec{A}^2$ is proportional to laser intensity $I$. Usually, the laser intensity is taken to be a Gaussian distribution; however, in the limits explored in this paper the corrections are required \cite{Esarey:95}. These limits are the time that the laser is pumping the system $\tau_0=100$fs and the wavelength $\lambda=800$nm. Now, the important terms in the equation above in the pump-on regime are the time-dependent terms on the right hand side. Thus, the equations reduce to
\begin{equation}
\begin{split}
    -\frac{1}{\Gamma_1}\partial_t \bar{\psi}_1 &=    b_1  e^2 \vec{A}(t)^2 \bar{\psi}_1,
    \\ -\frac{1}{\Gamma_2}\partial_t \bar{\psi}_2 &=   (-\alpha_2+ 4 b_2  e^2 \vec{A}(t)^2)\bar{\psi}_2.
     \end{split}
     \label{appc2}
\end{equation}

The mean-field solutions for $t \leq \tau_0$ are
\begin{equation}
    \begin{split}
        \bar{\psi}_1(t) &=\bar{\psi}_1^0  \exp \left(-\Gamma_1 b_1 I_0 \frac{\tau_0}{4}\zeta(t)\right),
        \\ \bar{\psi}_2(t) & = \bar{\psi}_2^0 \exp \left(-\Gamma_2\alpha'_2 t\right) \exp \left(-\Gamma_2 b_2 I_0 \tau_0\zeta(t)\right).
    \end{split}
    \label{appc3}
\end{equation}
where $\zeta(t)=\sqrt{2\pi}\left(1+ \frac{1}{\sigma^2}\right)\text{Erf}\left(\frac{ \sqrt{2}t} {\tau _0}\right)- 4\frac{1}{\sigma^2}\frac{t}{\tau_0} e^{-2(\frac{t}{\tau_0})^2}$.

\subsection{Pump-off regime, early times}
The mean-field solutions for $\tau_0 \leq t $ are
\begin{equation}
    \bar{\psi}_2(t)  = \bar{\psi}_2^0 e^{\Gamma_2\left(\alpha_2 t -(\alpha'_2 + \alpha_2)\tau_0+ (\alpha'_2+ \alpha_2)t_c (e^{-\frac{t-\tau_0}{tc}}-1)-b_2 I_0 \tau_0\zeta(\tau_0)\right)}.
    \label{appc4}
\end{equation}

If we consider the CDW order to follow
\begin{equation}
    -\frac{1}{\Gamma_1} \partial _t \bar{\psi}_1=(-\alpha_1 + 4 u_1 \bar{\psi}_1^2 + 2 c \bar{\psi}_2^2)\bar{\psi}_1,
    \label{appc5}
\end{equation}
for a general expression of $\bar{\psi}_2(t)$, the equation describing $\bar{\psi}_1(t)$ becomes
\begin{equation}
    \bar{\psi}_1(t) = \frac{\bar{\psi}_1(\tau_0)e^ { \int_{\tau_0}^{t} \left(\alpha_1 \Gamma_1 - 2 c \Gamma_1 \bar{\psi}_2(t')\right) dt' }}{\sqrt{1 +8 \Gamma_1 u_1 \bar{\psi}_1^2(\tau_0)  \int_{\tau_0}^{t} e^ {2 \int_{\tau_0}^{t'} (\alpha_1 \Gamma_1 - 2 c \Gamma_1 \bar{\psi}_2(\Tilde{t})) d\Tilde{t}}  dt'}}.
\label{appc6}
\end{equation}

For $\tau_0\leq t \leq t_c$ the equations can be approximated as,
\begin{equation}
    \bar{\psi}_2(t)=\bar{\psi}_0 e^{-\Gamma_2b_2 I_0 \tau_0\zeta(\tau_0)} e^{-\alpha'_2 \Gamma_2 t},
    \label{appc7}
\end{equation}
where we define $g=\bar{\psi}_2^0 e^{-\Gamma_2b_2 I_0 \tau_0\zeta(\tau_0)}$, yielding
\begin{equation}
    \bar{\psi}_1(t)=\bar{\psi}_1(\tau_0) \frac{e^{\Gamma_1 \left( \alpha_1 t + \frac{g c e^{-2 \alpha '_2 \Gamma_2 t}}{\alpha '_2 \Gamma_2} \right )}}{\sqrt{\exp{\left( \frac{2 \Gamma_1}{\alpha'_2 \bar{\Gamma_2}} \left( \alpha'_2 \alpha_1 \tau_0 + c g e^{-2 \alpha'_2\Gamma_2\tau_0}\right)\right ) + 8 \Gamma_1 u_1 \bar{\psi}_1^2(\tau_0)\int_{\tau_0}^{t} O(t') \,dt'}} },
    \label{appc8}
\end{equation}
where $\bar{\psi}_1(\tau_0)$ is determined from Eq.\eqref{appc3} at $t=\tau_0$ and $O(t') = \exp\left ( 2 \Gamma_1 \right( \alpha_1 t' + \frac{c g}{\alpha'_2 \Gamma_2} e^{-2 \alpha'_2 t'} \left) \right )$. 

\subsection{Metastability dynamics}
Using the Taylor expansion, we calculate the time at which the second order $\bar{\psi}_2$ becomes 1/p-th of its equilibrium value. This value $\bar{\psi}_{2,p}$ should be $\bar{\psi}_{2,p} \ll \bar{\psi}_1^0$, with
\begin{equation}
    t_p = \frac{1}{\alpha'_2}\left ( \frac{\ln p}{\Gamma_2} - b_2 I_0 \frac{\tau_0}{4}\zeta(\tau)\right ).
    \label{appc9}
\end{equation}

Let us consider Eq.\eqref{appc5} but without the $\psi_2$-dependence. This would be a good approximation for $t\geq t_p$ This would give us
\begin{equation}
    \bar{\psi}_1(t) = \sqrt{\frac{\alpha_1}{4 u_1 -\frac{4 u_1 \bar{\psi}^2_{1,p}-\alpha_1}{\bar{\psi}^2_{1,p}} e^{-2 \Gamma_1 \alpha_1 (t-t_p)}}}.
    \label{appc10}
\end{equation}
$\psi_{1,p}$ can be calculated from Eq.\eqref{appc6} at $t=t_p$. It is worth mentioning that there is another way to determine this time. One can simply look at the time when $\alpha_2(t)<0$, $t=t_c \ln(\frac{\alpha_2'}{\alpha_2}+1)+\tau_0$. Now, if we include the first-order $\psi_2$-dependence and approximate $\psi_2$. 
\begin{equation}
    -\frac{1}{\Gamma_2} \partial _t \bar{\psi}_2=(\alpha_2(t)+ 2 c \bar{\psi}_1^2)\bar{\psi}_2,
    \label{appc11}
\end{equation}
\begin{equation}
    \bar{\psi}_2(t)=\bar{\psi}_{2,p}\left( \frac{4 u_1 e^{2 \Gamma_1 \alpha_1(t-t_p)}-\frac{4 u_1 \bar{\psi}^2_{1,p}-\alpha_1}{\bar{\psi}^2_{1,p}}}{4 u_1-\frac{4 u_1 \bar{\psi}^2_{1,p}-\alpha_1}{\bar{\psi}^2_{1,p}}}\right)^{-\frac{\Gamma_2}{\Gamma_1}\cdot\frac{c}{4 u_1}}e^{\Gamma_2\left(t_c(\alpha'_2+\alpha_2)\left(e^{-\frac{t-\tau_0}{tc}}-e^{-\frac{t_p-\tau_0}{tc}}\right)+\alpha_2(t-t_p)\right)},
    \label{appc12}
\end{equation}
We get a solution that is valid in the range of $t_p<t<t_r$ where $t_r$ is determined by $\bar{\psi}_2(t_r)=\bar{\psi}_1^0=\bar{\psi}_{2,r}$. This encourages us to calculate the lifetime of the metastable state. Let's calculate the time that the second order is smaller and equal to the first order by setting Eq.\eqref{appc12} equal to $\bar{\psi}_1^0$. Therefore,
\begin{equation}
    \frac{2 u_1}{\Gamma_2 c} \ln\frac{\bar{\psi}_{2,p}}{\bar{\psi}_1^0}=\left(\alpha_1-2\frac{u_1}{c}\alpha_2\right)(t_r-t_p)- \frac{2 u_1}{c}t_c(\alpha_2'+\alpha_2)e^{-\frac{t_p-\tau_0}{t_c}}(e^{-\frac{t_r-t_p}{t_c}}-1).
    \label{appc13}
\end{equation}

Equation \eqref{appc13} shows the implicit solution to the metastability time, $t_r$. For small $t_r$ it can be approximated as $t_r-t_p=\frac{2u_1}{\Gamma_2(\alpha_1 c-2 u_1 \alpha_2)}\ln(\frac{\bar{\psi}_{2,p}}{\bar{\psi}_1^0})$. This approximation predicts the metastability lifetime accurately to the order of magnitude. An important point to note here is the solutions in Eqs.\eqref{appc10}-\eqref{appc13} are more accurate in the large $u_1$ limit since we are ignoring the effect of $\bar{\psi}_2$ in the dynamics of $\psi_1$. The quartic term in the energy functional of $\psi_1$ should be larger than the competing term for accurate results. 
\section{Correlations}
As discussed in the main text, the correlations lengthen the time that the system spends in the local energy minimum and also renormalize the energy landscape. However, the effect of the cross-correlation $C^{ij}_\mathbf{k}$ needs a more careful treatment as this term was not studied previously in the literature.

\begin{figure}[h]
    \centering \includegraphics[width=0.5\linewidth]{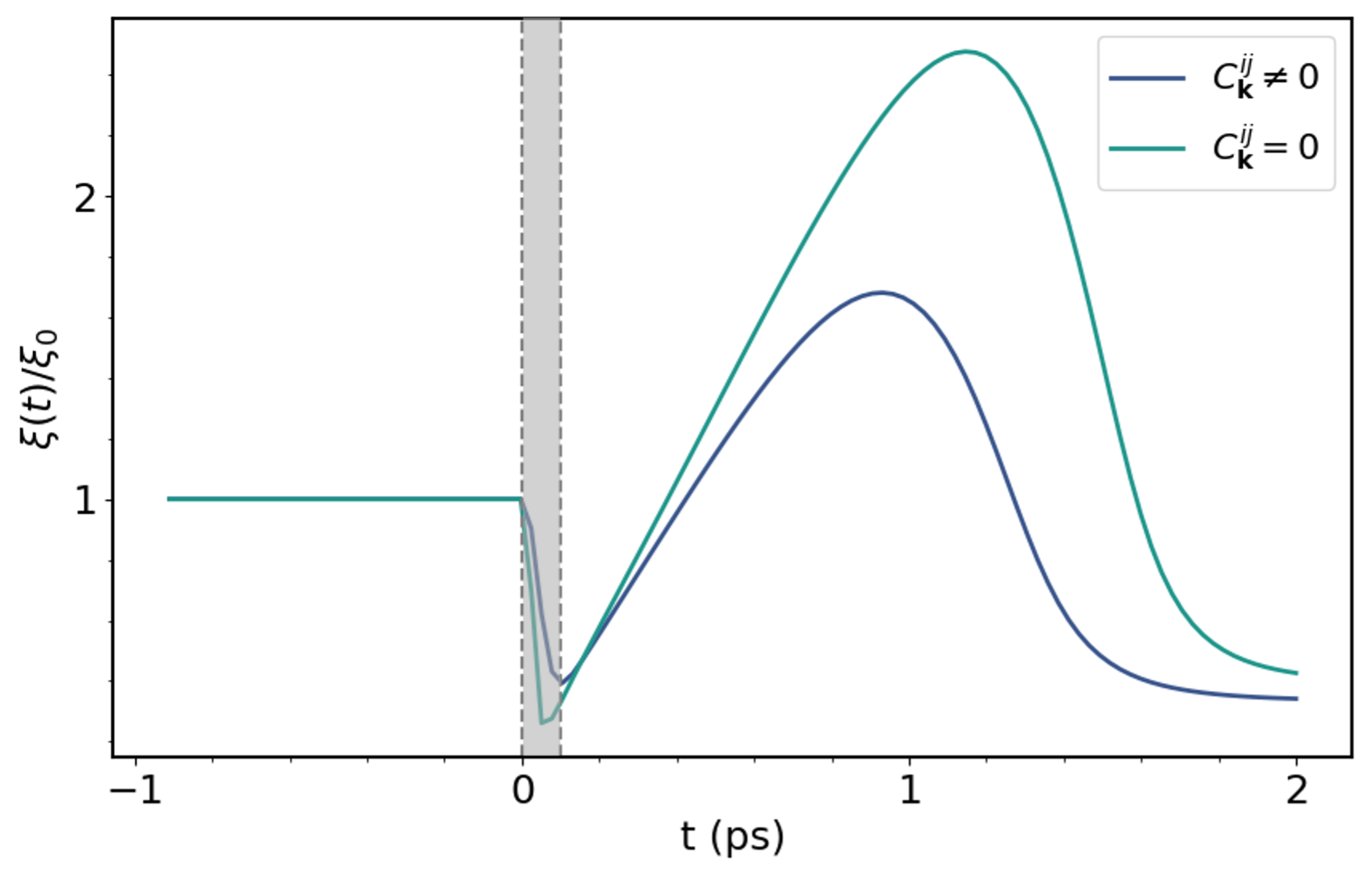}
    \caption{\footnotesize{The teal line shows the correlation length of the primary order, $\psi_1$, without cross-correlation, displaying a much larger increase compared to the blue line, where cross-correlation is included. This reflects limited energy exchange between the two orders, resulting in more robust orders, independent of metastability. The values are: $\alpha_\mathrm{CDW}=1.0$, $\alpha_\mathrm{SC}=1.1$, $\alpha_\mathrm{SC}'=3.5$, $b_\mathrm{CDW}=7.0$, $b_\mathrm{SC}=2.0$, $u_\mathrm{CDW}=2.0$, $u_\mathrm{SC}=1.0$, $c=1.5$, $\Gamma_\mathrm{CDW}=2.5$ps\textsuperscript{-1} $\Gamma_\mathrm{SC}=1.0$ps\textsuperscript{-1}, $e^2I_0^2=1.0$, and $t_c=1.0$ ps. }}
    \label{fig5}
\end{figure}

If we take the meanfield values of each order to be positive then according to Table~\ref{tab:table_regimes}, $C^{ij}_\mathbf{k}<0$, and this would make the recovery time faster in the dynamical trapping state. Now, if we take a look at \hyperref[fig5]{Fig.~\ref*{fig5}} we can see that in the absence of this term in the numerical calculations the orders exhibit a larger correlation length and this agrees with the previous argument of having faster recovery, more robust orders. Also, it is important to note that the Gaussian noise $\eta$ coming from the bath has no direct effect on the dynamics of this term and this term purely comes from the competition term $c\psi_1\psi_2$ and has a nonthermal nature.  

\end{document}